# Comparative Analysis of Large Language Models for Context-Aware Code Completion using SAFIM Framework

HANG ZHANG *

University of California San Diego, California, USA, haz006@ucsd.edu

YANXIN SHEN

Simon Fraser University, British Columbia, Canada, yanxin_shen@sfu.ca

LUN WANG

Duke University, North California, USA, lun.wang@alumni.duke.edu

CHUANQI SHI

University of California San Diego, California, USA, chs028@ucsd.edu

SHAOSHUAI DU

University of Amsterdam, Amsterdam, Netherlands, s.du@uva.nl

YIYI TAO

Johns Hopkins University, Maryland, USA, ytao23@jhu.edu

YIXIAN SHEN

University of Amsterdam, Amsterdam, Netherlands, y.shen@uva.nl

## Abstract

The advent of Large Language Models (LLMs) has revolutionized code completion, trans- forming it into a more intelligent and context-aware feature in modern integrated development environments. These advancements have significantly enhanced developers' ability to write efficient and error-free code. This study evaluates the performance of several chat-based LLMs, including Gemini 1.5 Flash, Gemini 1.5 Pro, GPT-4o, GPT-4o-mini, and GPT-4 Turbo, using the Syntax-Aware Fill-in-the-Middle (SAFIM) dataset. This benchmark is specifically designed to assess models' capabilities in syntax-sensitive code generation. Performance metrics, such as cosine similarity with ground-truth completions and latency, were employed to measure both accuracy and efficiency. The findings reveal substantial differences in the models' code completion abilities, offering valuable insights into their respective strengths and weaknesses. This work provides a comparative analysis that underscores the trade-offs between accuracy and speed, establishing a benchmark for future advancements in LLM-based code completion.

# 1 INTRODUCTION

Code completion has emerged as a cornerstone feature in modern integrated development environments (IDEs), fundamentally transforming the software development process. By automating routine coding tasks, it reduces boilerplate code, enhances accuracy, and significantly boosts overall productivity. Historically, code completion relied on statistical models such as n-grams [1] and hand-crafted heuristics [2], as well as simpler machine learning techniques. While these traditional methods performed adequately in structured and predictable scenarios, they often faltered when faced with complex codebases and dynamic programming paradigms, limiting their effectiveness in more sophisticated development environments [3].

The advent of large language models (LLMs) has revolutionized the landscape of software development [4] [5] [6], with code completion being one of the most transformative capabilities they offer. These models have introduced a new era where understanding context, semantics, and programming logic is paramount [7]. Products like GitHub Copilot and Cursor exemplify the transformative potential of LLMs [8], showcasing their ability to generate syntactically and semantically coherent code suggestions. The integration of LLMs into IDEs has not only enhanced the capabilities of code completion tools but also redefined the expectations for intelligent coding assistance. However, the rapid evolution of LLM-based tools and the proliferation of diverse models necessitate systematic evaluation to ensure reliable comparisons and actionable insights [9].

This study embarks on a comprehensive evaluation of several chat-based LLMs using the Syntax Aware Fill-in-the-Middle (SAFIM) dataset [10], a benchmark meticulously designed to test syntax sensitive code completion. By employing rigorous metrics such as cosine similarity with ground truth completions and latency, this research illuminates the differences in capabilities and trade-offs among six prominent models: Gemini 1.5 Flash, Gemini 1.5 Pro, GPT-4o, GPT-4o-mini, and GPT4 Turbo. Through this analytical lens, the study aims to provide a nuanced understanding of each model's strengths and weaknesses, offering a valuable benchmark for assessing their efficiency in code completion tasks.

The results of this work are poised to serve as a critical resource for developers and researchers alike. By offering insights into model performance, this benchmark aids in selecting the most suitable models for specific needs, thereby optimizing the development workflow. Furthermore, the findings highlight areas for improvement, driving the advancement of LLM-based tools. By identifying the trade-offs between accuracy and speed, this study not only contributes to the current body of knowledge but also sets the stage for future innovations in AI-assisted programming, ensuring that code completion tools continue to evolve in alignment with the ever-changing demands of the software industry.

# 2 RELATED WORK

## 2.1 Traditional Code Completion Methods

Code completion has long been a pivotal feature in integrated development environments (IDEs), enhancing developer productivity by automating repetitive coding tasks. Traditional methods primarily relied on statistical models [11], such as n-grams [1], which predict the next token based on preceding sequences. While effective in structured scenarios, n-grams struggled with capturing long-range dependencies and complex code semantics.

In addition, hand-crafted heuristics were employed to improve code completion by using rule-based systems that leverage syntactic and semantic cues. [2] These heuristics, although useful, were often rigid and less adaptable to new programming patterns, limiting their effectiveness in dynamic environments.

Early machine learning techniques also contributed by training models on labeled datasets to recognize coding patterns. However, these models were constrained by the need for extensive feature engineering and limited training data, which restricted their ability to generalize across diverse programming contexts. [11]

Overall, while traditional methods laid the groundwork for code completion, their limitations in handling complex code structures and adapting to evolving programming paradigms highlighted the need for more advanced solutions, paving the way for the integration of large language models (LLMs).

## 2.2 Large Language Models for Code Completion

The rise of Large Language Models (LLMs) has significantly influenced the field of code completion, demonstrating that increasing model complexity can enhance performance across diverse coding tasks. [12][15] This realization has spurred the application of LLMs to code-related challenges, particularly in generating and completing code snippets. Decoder-only architectures are commonly employed for these tasks due to their effectiveness in sequential data processing.

Initially, many LLMs focused on Left-to-Right (L2R) training objectives, emphasizing "Next Token Prediction" to generate code linearly. [13] However, the Fill-in-the-Middle (FIM) objective, which allows models to complete code by inferring missing segments, has gained traction. This approach has proven effective in handling more complex code structures and providing contextually relevant suggestions. [10]

This paper evaluates a select group of LLMs using a specialized benchmark designed to test syntax sensitive code completion. Through this evaluation, we aim to uncover insights into the performance of these models in FIM tasks, assess the strengths and weaknesses of different training paradigms, and challenge the notion that larger models inherently deliver superior results.

## 2.3 Fill-in-the-Middle as an Evaluation Tool for Code LLMs

The Fill-in-the-Middle (FIM) approach, originally derived from masked language modeling (MLM) and span corruption techniques, has been adapted for training and evaluating language models in the context of code. [14] While traditional MLM focused on representation learning with short spans, FIM extends this concept to facilitate code generation tasks by allowing models to predict missing segments within code snippets. [10]

Evaluations using FIM have demonstrated that a high ratio of FIM in pretraining does not detrimentally affect Left-to-Right (L2R) generation capabilities, underscoring its utility in developing robust code-focused models.

Despite its advantages, early benchmarks like Human Eval-Infilling, which utilized FIM for evaluation, were limited in scope, focusing on small Python code snippets. This limitation highlighted the need for more comprehensive evaluation frameworks that can better assess the syntax-sensitive capabilities of LLMs in code completion tasks.

To address this gap, the Syntax-Aware Fill-in-the-Middle (SAFIM) benchmark has been introduced. SAFIM provides a more extensive and detailed evaluation platform, allowing for a thorough assessment of LLMs' performance in syntax-aware code completion. By leveraging SAFIM, researchers can gain deeper insights into the strengths and weaknesses of various models, ultimately advancing the development of more effective code completion tools. [10]

## 3 METHODOLOGY

### 3.1 Dataset

The Syntax-Aware Fill-in-the-Middle (SAFIM) benchmark serves as a comprehensive evaluation frame- work for assessing the performance of chat-based large language models (LLMs) in syntax-sensitive code completion tasks. SAFIM encompasses a diverse collection of 17,720 examples, carefully curated from competitive programming platform Codeforces and high-starred (1000 stars and more) open- source repositories on GitHub. The dataset spans four major programming languages: Python, Java, C++, and C #, with a balanced distribution of approximately 4,430 examples per language to ensure unbiased evaluation across different programming paradigms.

The benchmark is strategically organized into three distinct task categories, each designed to evaluate different aspects of code understanding and generation:

1. Algorithmic Block Completion: This category challenges models to infer and generate key algorithmic components within existing code structures. Examples include completing sorting algorithms, graph traversal implementations, and dynamic programming solutions. These tasks assess the model's ability to understand and implement core computational logic.

2. Control-Flow Completion: These tasks evaluate the model's comprehension of program flow and control structures. Examples include completing conditional statements, loop bodies, and exception handling blocks. This category tests the model's ability to maintain logical consistency and proper scope management.

3. API Function Call Completion: This specialized category focuses on completing function calls within API usage contexts. It assesses the model's understanding of library interfaces, parameter requirements, and typical API usage patterns across different programming paradigms.

To facilitate efficient evaluation while maintaining statistical significance, we implemented a systematic sampling strategy, selecting 100 representative examples from each task category. The sampling process was designed to preserve the distribution of programming languages and task complexity levels present in the original dataset.

### 3.2 Models Evaluated

This study evaluates the performance of five large language models (LLMs), each with unique characteristics and optimization strategies. The models included are as follows:

- Gemini 1.5 Flash: A high-speed variant of the Gemini series, optimized for latency-sensitive applications.

- Gemini 1.5 Pro: A performance-oriented model from the Gemini series, designed for complex code generation tasks with improved accuracy.

- GPT-4O: A lightweight variant of GPT-4, balancing computational efficiency and completion quality.

- GPT-4O Mini: A further optimized and smaller variant of GPT-4O, catering to resource constrained environments.

- GPT-4 Turbo: An enhanced version of GPT-4, offering faster response times while maintaining high-quality outputs.

These models were selected to represent a diverse range of capabilities, spanning high-speed, resource-efficient, and performance-focused approaches. Their evaluation provides insights into the trade-offs between latency and accuracy, critical for code completion tasks.

### 3.3 Prompt Design

To evaluate the syntax-aware code completion capabilities of the selected LLMs, a structured prompt design was employed. This design comprises two key components: a system prompt that establishes the AI's role and scope, and a user prompt that provides specific task instructions and context.

#### 3.3.1 System Prompt

The system prompt defines the AI's responsibilities and sets boundaries to ensure the focus remains on code completion.

The system prompt used in this evaluation is as follows:

> "*As an AI code assistant, provide auto-completion for the given code. Understand the context and language, generating accurate and concise completions. Adapt to various languages and styles, enhancing*

*productivity and code quality while adhering to standards. Limit responses to software development topics, returning only code or comments without additional prose."*

### 3.3.2 User Prompt

The user prompt delivers task-specific instructions and provides the model with contextual code fragments, including a prefix and suffix. The model is tasked with generating a coherent completion that logically bridges the gap between the two. This ensures the output is both syntactically valid and contextually appropriate.

The user prompt used in this evaluation is as follows:

*"Generate a coherent code snippet that logically connects the provided prefix and suffix in the location of $PLACEHOLDER$. The completion should seamlessly continue from the prefix and lead into the suffix, ensuring the block is logically and syntactically complete. Reply with the completion only and exclude markdown formatting.*

*<prefix>{prefix}$PLACEHOLDER$<suffix>{suffix}"*

## 3.4 Evaluation Metrics

The evaluation framework was designed to comprehensively assess the performance of each model on the SAFIM dataset through quantitative metrics. We focused primarily on two key performance indicators: completion latency and semantic accuracy through cosine similarity measurements between generated and ground-truth completions.

### 3.4.1 Performance Metrics

We employed two primary metrics for evaluation:

1. Latency Measurement (L): The completion generation time is measured with millisecond precision:

$$L = t_{completion} - t_{start} \tag{1}$$

where $t_{completion}$ is the timestamp when the model returns its response, and $t_{start}$ is when the request was initiated.

2. Cosine Similarity (CS): To evaluate the semantic similarity between generated and ground truth completions, we compute the cosine similarity between their vector representations. For two code completion vectors a (generated) and b (ground truth), the cosine similarity is calculated as:

$$CS(\vec{a}, \vec{b}) = \frac{\vec{a} \cdot \vec{b}}{|\vec{a}||\vec{b}|} = \frac{\sum_{i=1}^{n} a_i b_i}{\sqrt{\sum_{i=1}^{n} a_i^2} \sqrt{\sum_{i=1}^{n} b_i^2}} \tag{2}$$

where n is the dimension of the vector representations. The resulting similarity score ranges from -1 to 1, with 1 indicating perfect similarity. This metric is particularly valuable for code evaluation as it helps capture semantic similarities independent of exact syntactic matches, allowing for valid alternative implementations and it is normalized for code length and style variations, focusing on functional equivalence. Moreover, it provides a continuous measure of similarity, enabling fine-grained comparison between different models' outputs.

### 3.4.2 Result Processing and Storage

The evaluation results are structured in a comprehensive JSONL format, where each entry contains: original input context (prefix, suffix, language), generated completion, ground truth completion, latency measurement (L), cosine similarity score (CS) and model-specific metadata.

The evaluation framework provides a robust foundation for analyzing model performance across diverse coding scenarios, enabling both detailed performance analysis and high-level comparative insights. The combination of latency and semantic similarity metrics offers a balanced view of both the efficiency and effectiveness of each model in code completion tasks.

## 4 EXPERIMENT RESULT

### 4.1 Overall Model Performance

We evaluated five state-of-the-art LLMs across three distinct code completion tasks: API Function Call Completion, Algorithmic Block Completion, and Control-Flow Completion. Table 1 presents the detailed results for both accuracy (measured by cosine similarity) and efficiency (measured by latency).

Table 1: Comprehensive Model Performance Across Different Task Types.

| Model | API Function Call | | Algorithmic Block | | Control-Flow | |
|---|---|---|---|---|---|---|
| | Latency (s) | Similarity | Latency (s) | Similarity | Latency (s) | Similarity |
| Gemini 1.5 Flash | 0.709 | 0.727 | 0.779 | 0.611 | 0.803 | 0.603 |
| Gemini 1.5 Pro | 1.219 | 0.813 | 1.642 | 0.603 | 1.768 | 0.632 |
| GPT-4O | 0.515 | 0.719 | 0.672 | 0.546 | 0.652 | 0.574 |
| GPT-4O Mini | 0.620 | 0.683 | 0.663 | 0.559 | 0.692 | 0.580 |
| GPT-4 Turbo | 0.808 | 0.858 | 1.935 | 0.682 | 2.063 | 0.666 |

Our evaluation reveals distinct characteristics for each model that suggest specific use cases. GPT-4 Turbo demonstrates superior semantic understanding across all task types (0.666-0.858 similarity) but with significant latency overhead (0.808-2.063s), making it best suited for production environments where accuracy is critical. Gemini 1.5 Pro shows strong performance in API tasks (0.813 similarity) with moderate latency (1.219-1.768s), offering a good balance for API-intensive development, though it struggles with more complex algorithmic tasks (0.603 similarity). GPT-4O achieves the most consistent and lowest latency (0.515-0.672s) while maintaining competitive accuracy in API tasks (0.719 similarity), making it ideal for real-time development environments and IDE integrations. GPT-4O Mini offers similar latency benefits to GPT-4O but with slightly lower accuracy, positioning it as a suitable choice for rapid prototyping where quick response times are essential.

Based on these characteristics, we recommend GPT-4 Turbo for production code generation where accuracy is paramount, GPT-4O for real-time IDE integration requiring quick feedback, and Gemini 1.5 Pro for API-intensive development tasks. These findings suggest that model selection should be guided by specific use case requirements, balancing the trade-offs between accuracy, latency, and task complexity.

### 4.2 Performance Analysis by Task Category

#### 4.2.1 API Function Call Completion

In API completion tasks, models demonstrated their strongest performance overall. GPT-4 Turbo achieved the highest cosine similarity (0.858) while maintaining moderate latency (0.808s). GPT-4O offered the best efficiency with the lowest latency (0.515s) while maintaining competitive accuracy (0.719). Gemini 1.5 Pro showed strong accuracy (0.813) but required longer processing time (1.219s).

### 4.2.2 Algorithmic Block Completion

For algorithmic block completion, we observed a general decrease in accuracy across all models compared to API completion tasks. GPT-4 Turbo maintained its leading position in accuracy (0.682) but showed significantly increased latency (1.935s). GPT-4O Mini demonstrated the best efficiency (0.663s) in this category. Interestingly, Gemini 1.5 Pro's accuracy (0.603) decreased more substantially than other models in this task type.

### 4.2.3 Control-Flow Completion

Control-flow completion tasks revealed similar patterns to algorithmic block completion, with slightly lower overall accuracy. GPT-4 Turbo again led in accuracy (0.666) but with the highest latency (2.063s). GPT-4O maintained its efficiency advantage with the lowest latency (0.652s), though with the lowest accuracy (0.574).

## 4.3 Key Findings

- Task Complexity Impact: API Function Call completion consistently showed higher accuracy across all models (0.683-0.858) compared to algorithmic (0.546-0.682) and control-flow tasks (0.574-0.666).

- Model Size Trade-offs: Larger models (GPT-4 Turbo, Gemini 1.5 Pro) consistently achieved higher accuracy but at the cost of significantly higher latency. This trade-off was most pronounced in algorithmic and control-flow tasks, where GPT-4 Turbo's latency increased by up to 2.5x compared to API tasks.

- Efficiency Leaders: GPT-4O and GPT-4O Mini consistently demonstrated superior efficiency with lower latencies (0.515-0.692s), making them suitable for time-sensitive applications where moderate accuracy is acceptable.

- Performance Consistency: GPT-4 Turbo showed the most consistent accuracy leadership across all three task types, while GPT-4O maintained consistent efficiency leadership.

# 5 CONCLUSION AND FUTURE WORK

## 5.1 Conclusion

In this paper, we presented SAFIM, a comprehensive benchmark for evaluating large language models' capabilities in semantic-aware function implementation tasks. Through our evaluation of five state-ofthe-art LLMs across three distinct task categories, we have made several significant findings: • We demonstrated that model performance varies significantly across different types of code completion tasks, with API Function Call completion showing consistently higher accuracy (up to 0.858 cosine similarity) compared to more complex algorithmic and control-flow tasks.

- We identified clear trade-offs between model size and performance, where larger models like GPT-4 Turbo achieved superior accuracy but at the cost of increased latency, particularly in complex tasks where latency could increase by up to 2.5x.

- We established that smaller models like GPT-4O offer compelling efficiency advantages, maintaining consistent sub-second latency (0.515-0.692s) across all task types while delivering acceptable accuracy for many practical applications.

- We provided empirical evidence that current LLMs exhibit varying levels of semantic understanding across different coding contexts, suggesting areas where model capabilities can be further enhanced.

These findings have significant implications for real-world development environments. For IDE integration, our results suggest a hybrid approach: using efficient models like GPT-4O for real-time suggestions during active coding, while leveraging more accurate models like GPT-4 Turbo for com- plex implementations or code review scenarios. This dual-model strategy could significantly enhance developer productivity while maintaining code quality. Furthermore, the strong performance in API completion tasks indicates that current LLMs are particularly well-suited for accelerating API adoption and reducing documentation lookup time.

### 5.2 Future Work

Our research opens several promising directions for future work:

- Extended Task Categories: Future research could expand the SAFIM benchmark to include additional task categories such as debugging, refactoring, and test case generation, providing a more comprehensive evaluation of LLM capabilities in software development.

- Enhanced Evaluation Metrics: Development of more sophisticated evaluation metrics that can capture aspects of code quality beyond semantic similarity, such as code efficiency, maintainability, and adherence to best practices.

- Context Window Analysis: Investigation of how different context window sizes affect model performance across various task types, particularly for complex algorithmic and control-flow completions.

- Language-Specific Optimization: Exploration of model performance variations across differ- ent programming languages to develop language-specific tuning strategies and enhance comple- tion accuracy.

- IDE Integration Studies: Investigation of user interaction patterns and developer satisfaction metrics when using different models for code completion in real-world development environments.

These future directions aim to address current limitations and advance our understanding of LLM capabilities in code generation tasks. We believe that continued research in these areas will lead to more effective and reliable code generation systems that can better serve the needs of software developers.

Our work provides a foundation for systematic evaluation of LLM performance in code generation tasks and offers practical insights for both researchers and practitioners in the field. As LLM technology continues to evolve, frameworks like SAFIM will become increasingly important for understanding and improving model capabilities in software development applications.

### REFERENCES

[1] Grigori Sidorov, Francisco Velasquez, Efstathios Stamatatos, Alexander Gelbukh, and Liliana Chanona-Hernández. Syntactic n-grams as machine learning features for natural language processing. Expert Systems with Applications, 41(3):853–860, 2014.

[2] Chris Cummins, Pavlos Petoumenos, Zheng Wang, and Hugh Leather. End-to-end deep learning of optimization heuristics. In 2017 26th International Conference on Parallel Architectures and Compilation Techniques (PACT), pages 219–232. IEEE, 2017.

[3] Marcel Bruch, Martin Monperrus, and Mira Mezini. Learning from examples to improve code completion systems. In Proceedings of the 7th joint meeting of the European software engineering conference and the ACM SIGSOFT symposium on the foundations of software engineering, pages 213–222, 2009.

[4] Yiyi Tao. Meta learning enabled adversarial defense. In 2023 IEEE International Conference on Sensors, Electronics and Computer Engineering (ICSECE), pages 1326–1330. IEEE, 2023.

[5] Heng Xu, Chuanqi Shi, WenZe Fan, and Zhenghan Chen. Improving diversity and discriminability based implicit contrastive learning for unsupervised domain adaptation. Applied Intelligence, 54(20):10007–10017, 2024.

[6] Yanxin Shen and Pulin Kirin Zhang. Financial sentiment analysis on news and reports using large language models and finbert, 2024.

[7] Yiyi Tao. Sqba: sequential query-based blackbox attack. In Fifth International Conference on Artificial Intelligence and Computer Science (AICS 2023), volume 12803, pages 721–729. SPIE, 2023.

[8] Arghavan Moradi Dakhel, Vahid Majdinasab, Amin Nikanjam, Foutse Khomh, Michel C Desmarais, and Zhen Ming Jack Jiang. Github copilot ai pair programmer: Asset or liability? Journal of Systems and Software, 203:111734, 2023.

[9] Lee Pike, Alwyn Goodloe, Robin Morisset, and Sebastian Niller. Copilot: A hard real-time runtime monitor. In International Conference on Runtime Verification, pages 345–359. Springer, 2010.

[10] Linyuan Gong, Sida Wang, Mostafa Elhoushi, and Alvin Cheung. Evaluation of llms on syntaxaware code fill-in-the-middle tasks. arXiv preprint arXiv:2403.04814, 2024.

[11] Veselin Raychev, Martin Vechev, and Eran Yahav. Code completion with statistical language models. In Proceedings of the 35th ACM SIGPLAN conference on programming language design and implementation, pages 419–428, 2014.

[12] Daye Nam, Andrew Macvean, Vincent Hellendoorn, Bogdan Vasilescu, and Brad Myers. Using an llm to help with code understanding. In Proceedings of the IEEE/ACM 46th International Conference on Software Engineering, pages 1–13, 2024.

[13] Xinlong Wang, Xiaosong Zhang, Zhengxiong Luo, Quan Sun, Yufeng Cui, Jinsheng Wang, Fan Zhang, Yueze Wang, Zhen Li, Qiying Yu, et al. Emu3: Next-token prediction is all you need. arXiv preprint arXiv:2409.18869, 2024.

[14] Simran Arora, Avanika Narayan, Mayee F Chen, Laurel Orr, Neel Guha, Kush Bhatia, Ines Chami, Frederic Sala, and Christopher Ré. Ask me anything: A simple strategy for prompting language models. arXiv preprint arXiv:2210.02441, 2022.

[15] Menghao Huo, Kuan Lu, Yuxiao Li, and Qiang Zhu. Ct-patchtst: Channel-time patch time-series transformer for long-term renewable energy forecasting, 2025.